\documentclass[12pt,a4paper,reqno]{article}

\usepackage{amsmath,latexsym,amssymb,amsthm}
\usepackage{graphicx}

\usepackage{amsmath}\usepackage{amssymb}
\usepackage{amsthm}

\newcommand{\be}{\begin{equation}}
\newcommand{\ee}{\end{equation}}

\newcommand{\F}{\overline{F}}

\renewcommand{\>}{\rangle}

\theoremstyle{definition}

\theoremstyle{remark}

\numberwithin{equation}{section}

\begin{document}

\title{\bf{Comment on ``Understanding the Frauchiger-Renner argument'' by Jeffrey Bub}}

\author{Anthony Sudbery$^1$\\[10pt] \small Department of Mathematics,
University of York, \\[-2pt] \small Heslington, York, England YO10 5DD\\
\small $^1$ tony.sudbery@york.ac.uk}

\date{2 September 2020}

\maketitle

\begin{abstract}

Bub has recently claimed that the Frauchiger-Renner argument does not require an assumption that the state vector undergoes projection after measurement. It is shown that this claim is not true, and an alternative understanding of the argument is offered.  

\end{abstract}


The argument \cite{FR2} for which Bub offers an understanding \cite{Bub:understandingFR} is intended to be a proof that the following three statements are inconsistent because they imply a contradiction.

Rule Q: If an agent $A$ has established that a quantum system $Q$ is in a state
$|\psi\>_Q$ at time $t_0$, and the Born probability of the outcome $\xi$ of a measurement of
an observable $x$ on $Q$ in the state $|\psi\>_Q$ completed at time $t$ is $1$, then agent $A$ can
conclude: ``I am certain that $x = \xi$ at time $t$.''

Rule C: If an agent $A$ has established: ``I am certain that another agent $A′$,
whose inferences about certainty are in accordance with Rules Q, C, and S, is certain
that $x = \xi$ at time $t$'', then agent $A$ can conclude: ``I am certain that $x = \xi$ at time $t$''.

Rule S: If an agent $A$ has established ``I am certain that $x = \xi$ at time $t$'', then
agent $A$ cannot also establish ``I am certain that $x \ne \xi$ at time $t$''.

The argument proceeds by considering a composite quantum system consisting of a quantum coin $C$, a qubit $Q$ and four agents $F$, $\overline{F}$, $W$, $\overline{W}$, which evolves in response to various actions by the agents (for details see \cite{FR2} and \cite{Bub:understandingFR}). Several authors (e.g. \cite{FRassumptions}) have pointed out that the argument relies on more assumptions than the three rules Q, C and S; in particular, that a measurement leads to projection of the quantum state to an eigenstate of the measured observable, at least for some observers \cite{MucinoOkun}. Bub denies this. He writes (in italics)

\begin{quote}\emph{The Frauchiger-Renner argument does not assume, nor needs to assume, anything other than a successive unitary evolution of the global quantum state through each stage of the experiment --- in particular, the argument does not assume that the quantum state undergoes a ‘collapse’ for observers inside a laboratory but not for outside observers.}\end{quote}

Bub's italics do not make this a true statement. Its falsity becomes clear in Bub's next paragraph but one\footnote{In a revised version (arXiv:2008.08538v2) of Bub's paper, this has become the next paragraph but two (the third paragraph on p.6). Bub has added a paragraph which is irrelevant to the point made in this note.}, where he describes the step in the Frauchiger-Renner thought experiment in which the agent $\overline{F}$ makes an inference about the result of a future measurement by $W$ of a variable $w$. This inference is based on the result of a measurement by $\F$ of the variable $r$ (the result of a quantum coin toss) and her subsequent action. To make the inference as required for the Frauchiger-Renner argument, $\F$ must apply unitary evolution (incorporating the future conduct of the experiment) to the state in which $r$ takes the value given by her measurement, \emph{not} to the global quantum state --- in other words, she must assume that the quantum state undergoes a projection (or ``collapse''). The outside observer $W$, on the other hand, does not assume this collapse.

It is so obvious that the argument requires this assumption of collapse that it is worth trying to understand how Bub could have thought it reasonable to deny it. This will bring out the true significance of the Frauchiger-Renner argument. Bub (following Frauchiger and Renner) takes it to be required by standard quantum mechanics that unitary evolution is to be applied to the state revealed by experiment; this is taken to be included in Rule Q. In his subsequent exposition Bub makes it clear that he regards this state as just one component $|\Psi(\text{tails})\>$ of the global state; apparently this is the basis for his claim that there is no assumption of collapse. But this conceals the assumption of collapse in the \emph{content} of the component $|\Psi(\text{tails})\>$, which includes $\F$'s mental state ``I am certain that $w =$ fail at $n:31$'', which she has inferred from the result of her measurement of $r$. Apparently he thinks that the evolution of each ``branch'' of the global state vector (in the terminology of Everettians) is independent of the other branches. The Frauchiger-Renner argument shows that this is not true in quantum mechanics.

How should $\F$ argue if she does not assume projection of the state vector after her measurement? Surely, if she observes the result $r =$ tails, she is correct to say ``I know that $r =$ tails'' and then to form certainties about the future on the basis of this fact? No. If she believes that there is a successive unitary evolution of the global quantum state, then she will say “I know that I am in a superposition and I will therefore apply quantum mechanics with this superposition as initial condition”.  In that case she does not believe that $W$ will observe $w =$ fail.

Does this mean that $\F$ doesn’t believe the evidence of her eyes? No. $\F$ sees that the quantum coin came down $r =$ tails. This is indeed a fact. It might have come down $r =$ heads, but it didn’t. But as a well-brought-up quantum physicist, she knows that although $r =$ tails is a fact, it is only a fact relative to her mental state, not a fact about the universal state. It is the universal state, with a component in which $r =$ heads, that determines the future evolution. This is the true lesson of the Frauchiger-Renner argument:

\begin{quote}  What might have happened, but didn’t, can still affect the future course of events. \end{quote}

This looks disastrous for the conduct of science. How can we know what might have happened, but didn't, without knowing the whole history of the universe? How can we proceed if we cannot trust our experiments to give us objective facts? These worries are resolved by considering the artificiality of the Frauchiger-Renner scenario, in which simple qubits are meant to represent sentient, reasoning agents. In reality, a system with sufficient complexity to represent such agents would have a macroscopic number of degrees of freedom and would necessarily interact with a macroscopic environment. For such systems the effect of decoherence \cite{Schlosshauer} is to make the effect of ``what might have happened but didn't'' utterly negligible, and to destroy the coherent superposition necessary for the Frauchiger-Renner argument. Nevertheless, the argument is valuable in demonstrating that parts of what we normally take to be standard quantum mechanics are, according to the strict theory, valid only (but reliably) for all practical purposes.


\end{document}